# Abundance and size-frequency distributions of boulders in Linné crater's ejecta (Moon)

Maurizio Pajola[1], Riccardo Pozzobon[2], Alice Lucchetti[1], Sandro Rossato[2], Emanuele Baratti[3], Valentina Galluzzi[4], Gabriele Cremonese[1]

1 INAF, Osservatorio Astronomico di Padova, Vicolo dell'Osservatorio 5, 35122, Padova, Italy.

2 Geosciences Department, University of Padova, Padova, Italy.

3 School of Civil Engineering, Department DICAM, University of Bologna, Bologna, Italy.

4 INAF-Istituto di Astrofisica e Planetologia Spaziali, Roma, Italy.

**Abstract**-------------------------------------------------------------------------------------

This paper presents the abundances and the size-frequency distributions (SFD) of the ejected boulders surrounding the Linnè crater, located on the Moon's Mare Serenitatis basin. By means of Lunar Reconnaissance Orbiter Camera high-resolution images we prepare a context geological map of the Linné crater as well as we identify ∼12000 boulders ≥ 4.4 m, with a maximum measured size of 30.8 m. The cumulative number of boulders per $km^2$ is fitted with a power-law curve with index -4.03 +0.09/-0.10. By studying the radial ejecta abundances, we find that the largest ones are located within the first 2 km from the crater's centre, while few tens of boulders with sizes < 8 m are detectable above 5 km from the crater's rim. We find that the Linné proximal ejecta blanket is slightly asymmetrical, as indicated in the geological map too, showing a density increase in the NE-SW direction. This may be the result of an oblique impact emplacement of the original impactor, or it may be explained with a perpendicular impact in the Mare Serenitatis location, but on a surface with lunar basalts with different local mechanical properties. By exploiting our boulders size density as a function of the distance from the crater's centre, we derive a possible regolith thickness at the Linnè impact of ~4.75 m, supporting similar values based on Earth-based radar and optical data in the Mare Serenitatis basin.

## 1.0 Introduction

The first boulders discovered on a non-terrestrial surface were observed on the Moon in 1965, by means of the Ranger probe photographs (Kuiper et al., 1965). Then, in 1977, the Viking spacecraft photographed the first Martian boulders (Binder et al., 1977), hence suggesting that they might be present on other solid planetary surfaces too. Nowadays, we know that boulders surrounding impact craters are present not only on the surface of the Earth (Senthil Kumar et al., 2014), Moon (Cintala and McBride, 1995) or Mars (Moore and Jakosky, 1989), but also on icy satellites (Martens et al., 2015), asteroids (Thomas et al., 2001; Michikami et al., 2008), as well as on the actively reshaping cometary surfaces (Pajola et al., 2015, 2016a, Mottola et al., 2015).

In the last decades, thanks to the increasing number of high-resolution images it became feasible to detect boulders, as well as to study their size-frequency distribution (SFD). By means of the boulder SFD analyses, it is possible to investigate a wide range of processes that occurred or are still occurring on a planetary/minor body surface (e.g., McGetchin et al., 1973; Garvin et al. 1981; Craddock et al. 2000; Ward et al. 2005; Grant et al. 2006; Yingst et al. 2007; Golombek et al. 2008). With the exception of planets like Mars, where erosive and depositional phenomena (e.g., Christensen P. R. 1986) generate or degrade boulders, or on active/sublimating cometary surfaces where cliff collapses (Pajola et al., 2017a) occur forming taluses, boulders are mainly considered the result of impact processes (Melosh 1989). They are therefore the remnant of the excavated rocky interiors showing the underlying mineralogical composition (Shoemaker 1987) and their SFD, as well as the maximum generated sizes, are directly related to the impactor composition and velocity, and to the impact site morphology and geological properties. Several craters surrounded by ejected boulders are indeed observed on asteroids (Geissler et al., 1996; Thomas et al., 2001; Michikami et al., 2008; Mazrouei et al. 2014; Kueppers et al., 2012) and on the Moon (Hartmann 1969; Shoemaker 1970; Cintala and McBride 1995; Bart and Melosh, 2010a). For the Moon's case, the study of the boulder SFD identified on different maria is a fundamental mean to study the distribution and layering thickness of the lunar volcanic basalts (Hiesinger et al., 2000; Weider et al., 2010). We then focus our attention on the Linné crater located in Mare Serenitatis (~2.2 km-diameter, Fig. 1), because it is considered one of the most recent impact craters formed on the surface of the Moon, showing a particularly fresh morphology[1] and surrounded by pristine ejecta (Basilevsky et al., 2013).

The aim of this work is therefore to study the fresh boulder SFD and how it changes with increasing distance and azimuth around the crater in order to infer the morphological structure of the pre-impacted basaltic site. In addition, by exploiting the Linné boulders size density, we aim to study the regolith thickness present in the west Mare Serenitatis basin.

The paper is structured as follows: after presenting the dataset and methodology, we will describe the geological setting of the Linné crater, providing a framework to the following boulder SFD analysis. We will then discuss the results, comparing them with other Moon studied craters, then contextualizing this work within the bigger lunar impact cratering picture. Implications on the impact event as well as on the regolith thickness of the Linné area will be presented too.

---

[1] Linné is a pristine crater that is characterised by an inverted truncated cone shape (Garvin et al., 2011).

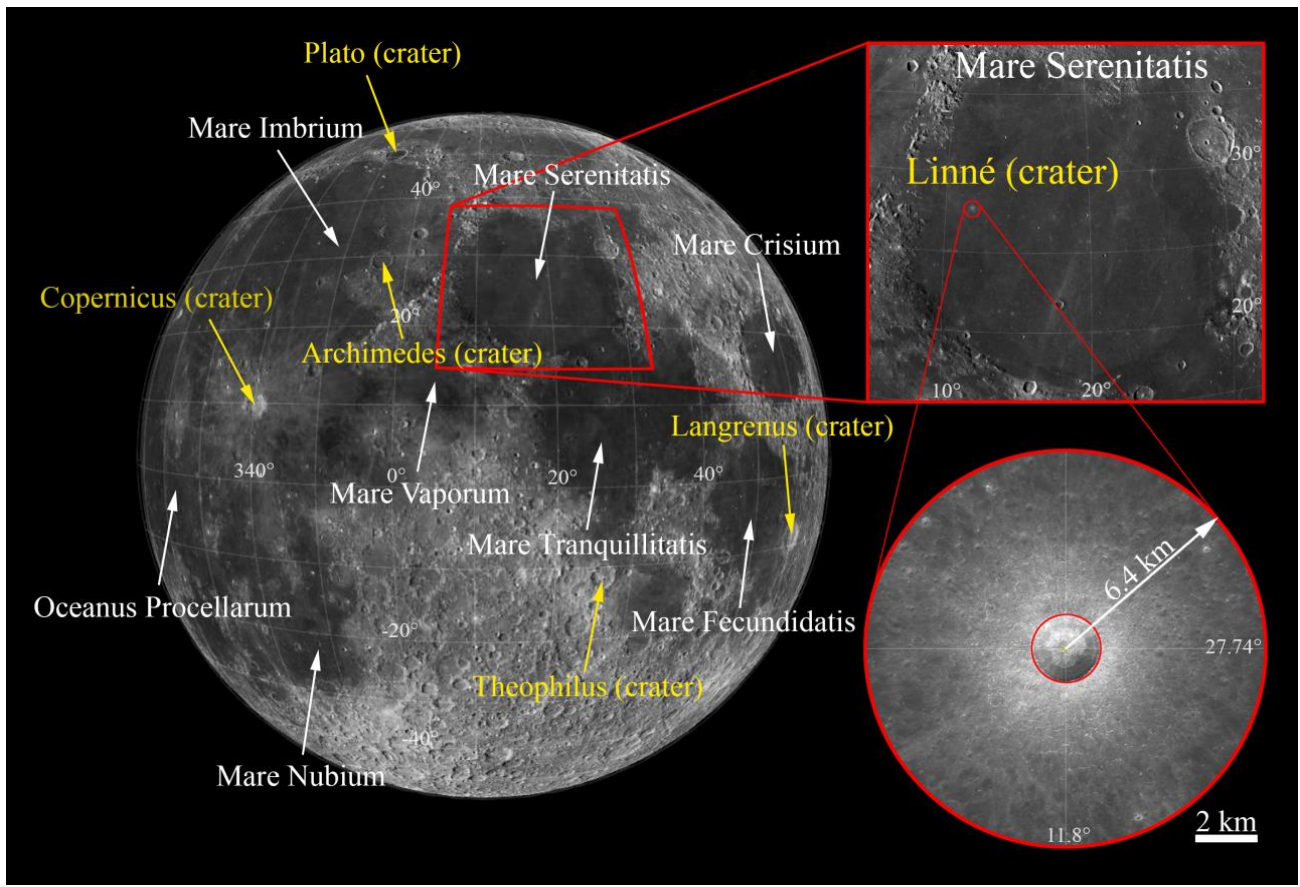

Fig. 1: Context image of the Moon with the locations of Mare Serenitatis and Linné crater, centred at latitude: 27.74°, longitude: 11.8°. The extension of 6.4 km from the crater's centre is explained in the 3.1 Geological mapping section.

## 2.0 Dataset and Methodology

### 2.1 Imagery and DTM

In order to study both the geological setting as well as the boulder distribution surrounding the Linné crater we downloaded two publicly available (http://wms.lroc.asu.edu/lroc/) Lunar Reconnaissance Orbiter Camera (LROC) mosaics made of two Narrow Angle Camera (NAC) images each (Klem et al., 2014). The first mosaic "NAC_ROI_LINNECTRHIA_E279N0118", hereafter called "HIA" image, is centred at 27.9°N latitude and 11.8°E longitude and it is characterised by a spatial scale of 1.10 m/pixel, see Fig. 2A. It was made through two NAC images both characterised by a solar incidence angle of 31° and a phase angle of 31-32°. The second mosaic "NAC_ROI_LINNECTRLOA_E282N0119", hereafter called "LOA" image, is centred at 28.2°N latitude and 11.9°E longitude and it is characterised by a spatial scale of 1.30 m/pixel. This was made through two NAC images characterised by a solar incidence angle of 51° and a phase angle of 50-52°.

In addition to the LROC image, in order to help the identification of geological units, their contacts, superimpositions and the identification of linear features, we downloaded from the same LROC site the Digital Elevation Model (DEM) called "NAC_DTM_LINNECRATER_E280N0120". This DEM completely covers the Linné crater at 2.0 m ground sampling distance (see Fig. 2B), while it partially covers its surrounding distal ejecta deposits (Fig. 5).

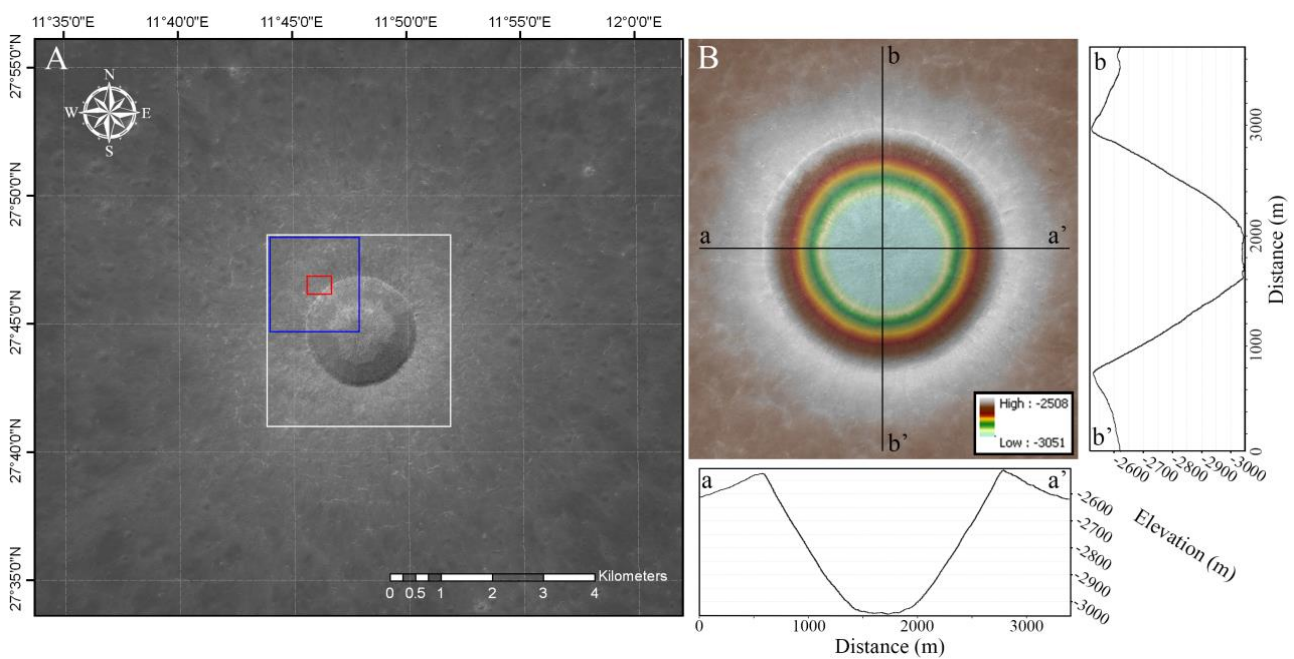

Fig. 2: A) A section of the LROC NAC image covering the Linné crater used for our boulder and geological analyses. The white box shows the extension of Fig. 2B, the blue box shows the extension of Fig. 3A-D, while the red box shows the extension of Fig. 4. B) The LROC generated DEM superimposed in transparency on the NAC image. The elevation bar indicated is in metres. The vertical profiles *aa'* and *bb'* are also presented.

## 2.2 The Geological Mapping

In order to identify the principal geological units that characterise the Linné crater, we used both the "HIA" and the "LOA" images. On average, the "HIA" frame (Fig. 3A), with a spatial scale of 1.1 m/pixel, provides uniform illumination conditions that enhance brightness variations with a SE-NW illumination direction. However, it was often necessary to adjust the gamma value of the image (i.e. grey contrast) to investigate those areas that were too dark or too bright in the frame. On the other hand, the "LOA" frame (Fig. 3B) presents a contrast that is too high for identifying clear terrain boundaries, with a spatial scale of 1.3 m/pixel; however, it provides a sharper view of the ridges found on the Linné ejecta blanket. Moreover, it provides an opposite illumination direction with respect to the former frame, which permits to better investigate the southeastern inner part of the crater. The NAC DTM does not provide a full coverage of the area; hence it is not useful for obtaining a uniform interpretation of the morphology all around the crater, yet it was used to support the selection of the crater main units (Fig. 3C). By investigating these datasets, it was possible to map the geological setting of Linné crater by using a mapping scale of 1:2,000 to 1:5,000. The geological contacts were divided into a) certain, where brightness and morphology show unambiguous changes; b) approximate, where it was not possible to define a clear boundary between units, because of gradational or unclear changes in terrain type (Fig. 3D). Linear features were used to map the main and secondary crater rims, along with older buried crater rims. The concentric impact ridges and troughs found around the crater were mapped using the same symbology, since most of them alternate and the mapping scale does not permit to keep them distinct. A stereographic projection centred on the spherical centroid of the crater (i.e. 27.74°N; 11.8°E) was used to avoid geometric distortions as much as possible.

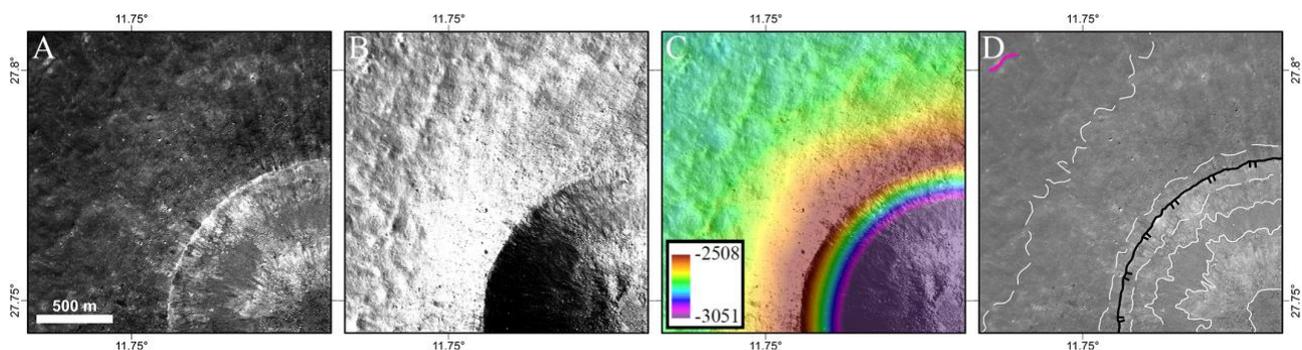

Fig. 3: Northwestern sector of Linné crater in stereographic projection centred at 27,7°N; 11.8°E as shown in: A) NAC_ROI_LINNECTRHIA_E279N0118 with enhanced grey contrasts; B) NAC_ROI_LINNECTRLOA_E282N0119; C) NAC_DTM_LINNECRATER_E280N0120 on top of B) with 50% transparency (as in Fig. 2B, the elevation values are in metres); D) same as A) but with unchanged grey contrasts with contacts and linear features on top. Solid white lines represent certain contacts; dashed white lines are approximate contacts; the black line is the crater rim with ticks toward the inner wall scarp; a short pink line is visible to the top left of the image representing a concentric impact ridge and trough system.

**2.3 Boulder identification**

For the ejected boulder identification, we mainly used the "HIA" image, given the better spatial scale and the lower phase angle with respect to the "LOA" image. Following the same approach of Pajola et al. (2015, 2016b, 2017b), we defined as a "boulder[2]" a positive relief detectable with the presence of an elongated shadow (if the phase angle is greater than 0°) and appearing "detached from the ground where it stands". The spatial scale of the "HIA" image is 1.1 m, hence by following the minimum three pixels sampling rule (Nyquist 1928), we can only detect ejected boulders with a minimum size of 3.3 m. Nevertheless, in order to obtain the boulder SFD statistics it is not uncommon to consider only boulders bigger than 4-7 pixels as a lower boundary. This is commonly done to provide a meaningful size-frequency statistic (Mazrouei et al., 2014; Pajola et al., 2017b), minimizing the likelihood of boulder misidentifications (Golombek and Rapp 1997, Golombek et al. 2003). The LROC image we used was taken with a phase angle of 31.5°, hence the presence of elongated shadows on the surface provided the possibility of identifying even smaller boulders (2 pixels, i.e. 2.2 m in size). However, in our statistics we considered the minimum diameter to be 4.4 m (i.e. 4 pixels), as the SFD below this value starts to roll over, indicating that the sampling might not be complete. Scarps and bedrock ledges are generally characterised by elongated and aligned shapes that cast shadows on the surface as well, but with contiguous, if not uninterrupted shapes that are different to those of the commonly isolated boulder' ones. Given the impact cratering origin of Linné, we suggest that all blocks observed surrounding the crater's rim are ejected boulders and not hills or mounds that typically emerge from the ground itself. Following these guidelines, we define all the identified features as "boulders".

---

[2] The official USGS size terms after Wentworth (1922) identifies as "boulders" all blocks that have diameters >0.25 m; therefore, in our case we can properly call all the identified blocks as boulders, given the minimum size of 4.4 m.

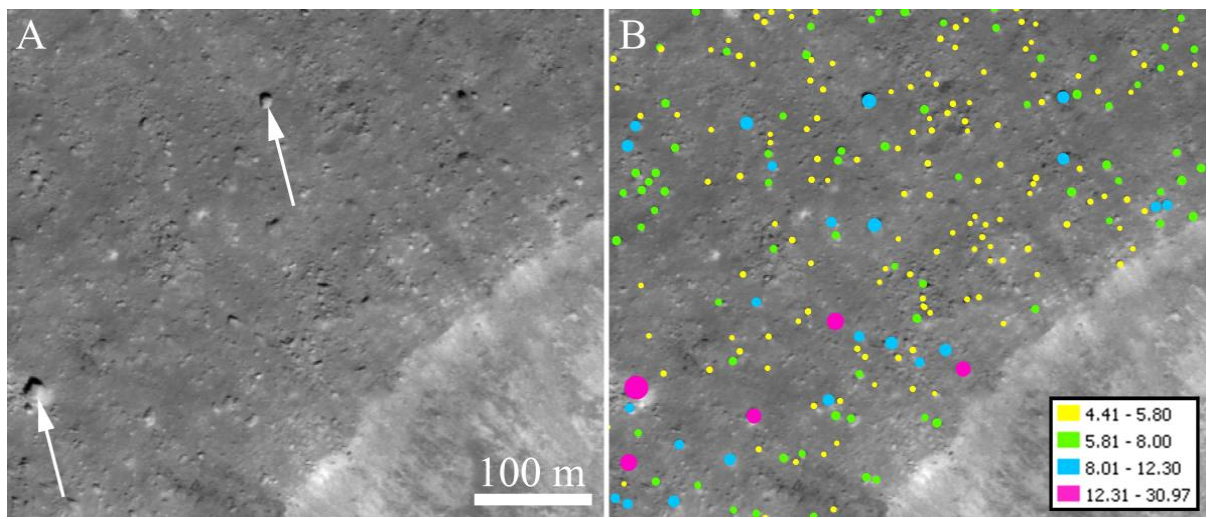

Fig. 4: Methodology used to identify the boulders located outside the Linné's rim. A) Subframe of the LROC "HIA" image used in our analysis. The white arrows indicate the direction of the sunlight. B) The detected boulders ≥ 4.4 m, grouped in size-categories (m). As it is possible to see, the illumination conditions (incidence of 31°, phase angle of 31.5°) provide the possibility to detect boulders with sizes of 2-3 pixels (2.2-3.3 m), but that are not considered in our statistics.

Fig. 4 shows the results of our boulder identification methodology in a test area. We used ESRI ArcGis 10.4 software for the manual boulder identification and mapping. As usually done in planetary sciences (Golombek et al., 2003; Michikami et al., 2008; Mazrouei et al. 2014, Kueppers et al., 2012, Pajola et al., 2016c,d) boulders were assumed to be circular in shape. We underline that despite using a circular shape for the considered boulders we are aware that they can have elongated shapes; nevertheless, in this work we are solely focusing on their maximum dimension distribution, and we do not consider the boulders morphometry. Hence, from the fitting circles we then derived the boulders' diameters. Consequently, in order to obtain the cumulative boulder size-frequency distribution per $km^2$, we used the corresponding area where we identified the ejecta. The data were grouped in bins of 1.1 m, i.e. the LROC "HIA" image resolution. Then, in all log-log plots, we fitted the data by using regression lines and we obtained the power-law index of each size distribution. The error bars for each bin indicate the root of the cumulative number of counting boulders divided by the considered area following Michikami et al. (2008).

## 3.0 Results

### 3.1 Geological mapping

The Linné crater is located at 27.74° latitude and 11.8° longitude in northwestern Mare Serenitatis, a basaltic smooth basin located on the nearside of the Moon (Fig. 1). It is a simple crater with a diameter *D* of 2.194 km, a depth *d* of 0.52 km and a *d/D* ratio of 0.237 (Garvin et al., 2011, Fig. 2). It likely formed in the late Copernican period, within the last 10 Ma (Stickle et al., 2016), and lays on top of a complex stratigraphic sequence of distinct volcanic events that piled up to form an homogeneous unit, as shown in the geological map reported by Hiesinger et al., 2000. The Mare Serenitatis stratigraphic sequence characterised by solidified lava flows could reach a few kilometers of depth from the maria surface, and each geological unit is made of lava flows several hundred metres thick (Pommerol et al., 2010; Weider et al., 2010).

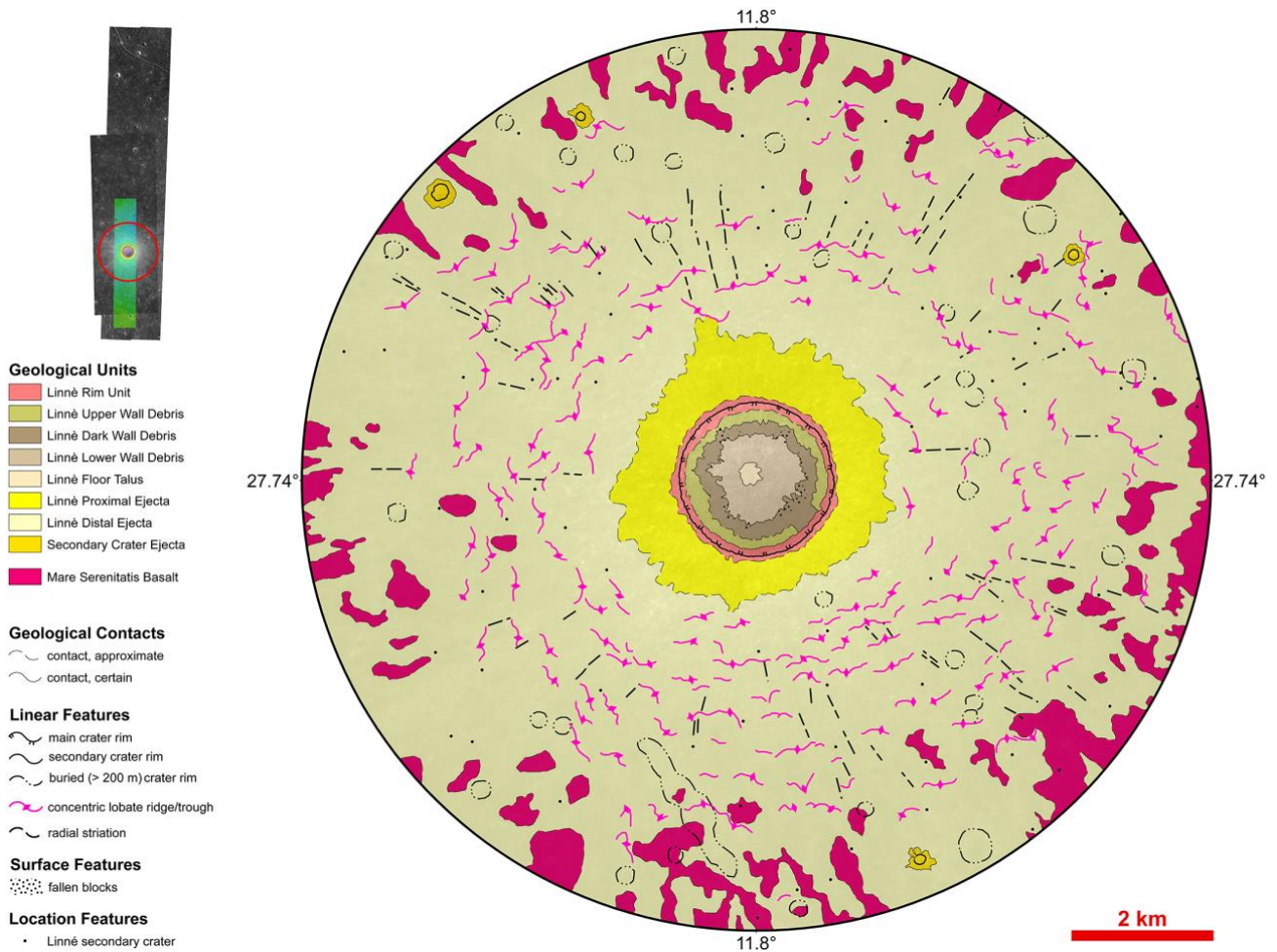

Fig. 5: Geological map of the Linné crater showing the different identified units. The non-interpreted version of this figure is in Fig. 1. Stereographic projection centred at 27.74°N; 11.8°E. Top-left: Extent of the NAC_ROI_LINNECTRHIA_E279N0118 image, the NAC_ROI_LINNECTRLOA_E282N0119 image, and the NAC_DTM_LINNECRATER_E280N0120 with the red circle indicating the extent of the geological map.

The geological map of the Linné crater encompasses nine units in total (Fig. 5). The bowl-shaped cavity of Linnè crater is mainly characterised by mass wasting deposits. These deposits were divided into four units based on their morphology and brightness. From top to bottom, the "Linné Upper Wall Debris" unit is characterised by medium to fine-grained, discontinuously bright wall debris; large blocks and gullies occur on top of this deposit. The "Linné Dark Wall Debris" unit is characterised by visibly darker deposits, with a coarser texture than the former unit. The "Linné Lower Wall

Debris" unit is characterised by coarse-grained deposits with abundant large blocks accumulating towards the floor. The small floor, decentered with respect to the spherical centroid defined by the crater rim, is solely characterised by the accumulation of the wall debris. All these deposits likely come from the "Linné Rim Unit", which is characterised by outcropping bedrock recognizable for visible fractured and tilted layers, mostly covered by ejected blocks and regolith. These layers likely pertain to the Mare Serenitatis basalt where the crater formed; however, they outcrop discontinuously and are better visible in the southern and southwestern part of the crater. Thus, only a thin ring of this unit was mapped in the geological survey, but it is expected to be found also farther from the rim where large ejected blocks make the morphology rougher. This area of continuous ejecta, covered in tens of metres large blocks –well recognizable in the "LOA" frame (e.g. Fig. 3B)– extends up to 1 km away from the crater rim and was mapped as "Linné Proximal Ejecta". Where the large blocks decrease in size and number, further away from the crater rim, a hummocky ejecta blanket extends for more than 3 km. This area of the crater is characterised by concentric bright systems of ridges and troughs generated by the impact wave, which are interconnected with radial and darker striations. All these features appear more pronounced in the "LOA" frame but only the most prominent ones were mapped, revealing a larger extent of ridges and troughs to the southeast and more striations to the northwest. This area covered in ridges was mapped as the "Linné Distal Ejecta" unit that becomes more and more discontinuous away from the crater centre, where the ejecta blanket becomes thinner and reveals dark window-patches of the underlying "Mare Serenitatis Basalt" unit. The main morphological features strictly related to the impact event are found within 6.4 km from the crater centre. Beyond this limit, the concentrical features are absent to the west of the crater and less frequent to the east. Sparse bright secondary craters are also found near this limit. The geological evidence suggests that, at this scale, impact related boulders should be no more recognizable beyond this limit and the selected extent is sufficient to provide context for the present study.

## 3.2 The Boulders SFD Analyses

### 3.2.1 The global boulder SFD analysis

For the global boulder SFD analysis, we considered a circular area with a radius of 5.3 km outside the craters' rim, i.e. 6.4 km from the crater's centre. As indicated in the geological section, beyond this distance the identifiable boulders are so few that a statistical analysis would be hardly meaningful. Over this area, 124.01 km$^2$ wide, we identified 46273 boulders $\geq$ 2.2 m, 12067 of which being $\geq$ 4.4 m. The corresponding density of boulders $\geq$ 4.4 m per km$^2$ is 137.85. Table 1 presents the statistics of the collected boulders, grouped in bins of 1.1 m.

Table 1: Statistics of the identified boulders, grouped per bin (1.1 m wide). The grey rows (size range 4.4-17.6 m) show the size range of the boulders considered for the power, exponential-law and Weibull fitting of Fig. 6.

| Bin* [m] | Absolute frequency | Cumulative number** | Cumulative number/km$^2$ |
|---|---|---|---|
| 2.2 | 10669 | 46273 | 528.60 |
| 3.3 | 19767 | 26506 | 302.79 |
| 4.4 | 14439 | 12067 | 137.85 |
| 5.5 | 6461 | 5606 | 64.04 |
| 6.6 | 2711 | 2895 | 33.07 |
| 7.7 | 1260 | 1635 | 18.68 |
| 8.8 | 680 | 955 | 10.91 |
| 9.9 | 371 | 584 | 6.67 |
| 11.0 | 188 | 396 | 4.52 |
| 12.1 | 116 | 280 | 3.20 |
| 13.2 | 86 | 194 | 2.22 |
| 14.3 | 71 | 123 | 1.41 |
| 15.4 | 35 | 88 | 1.01 |
| 16.5 | 26 | 62 | 0.71 |
| 17.6 | 18 | 44 | 0.50 |
| 18.7 | 15 | 29 | 0.33 |
| 19.8 | 7 | 22 | 0.25 |
| 20.9 | 7 | 15 | 0.17 |
| 22.0 | 5 | 10 | 0.11 |
| 23.1 | 2 | 8 | 0.09 |
| 24.2 | 1 | 7 | 0.08 |
| 25.3 | 1 | 6 | 0.07 |
| 26.4 | 2 | 4 | 0.05 |
| 27.5 | 3 | 1 | 0.01 |
| 28.6 | 0 | 1 | 0.01 |
| 29.7 | 0 | 1 | 0.01 |
| 30.8 | 1 | 1 | 0.01 |

\* the column reports the lower limit of each bin; \*\* the cumulative number is intended as the number of boulders with dimension equal or greater than the selected lower limit (i.e. first column).

In Fig. 6A we show the spatial distribution of all the identified ejected boulders ≥ 4.4 m surrounding the Linné crater, while in Fig. 6B we present the cumulative boulder number per km$^2$. The regression line used to interpolate the number of boulders per km$^2$ takes into account only those boulders that are between 4.4 m and 17.6 m in size. The choice to adopt 17.6 m as the upper limit is due to the size of the sample statistics, indeed, as shown in Table 1 such diameters correspond to the maximum limit where a similar cumulative number appears, hence indicating a poor statistics (Michikami et al., 2008, Pajola et al., 2015). On the other hand, the roll over above 17.6 m might be real and for this reason we decided to perform an exponential fit in the same size range. This fit seems to better reproduce the SFD trend for sizes larger than 14.3 m, but it overestimates the sizes in the range 7.7- 14.3 m, as well as it underestimates the sizes < 6.6 m, see Fig. 6B. A Weibull distribution, instead, seems to better fit the heavy-tailed distribution of the empirical data, particularly at the bigger sizes. On Earth, such distribution is largely used in fracture and fragmentation theory (Grady and Kipp, 1987; Brown and Wohletz, 1995; Turcotte, 1997), hence this could represent the real fragmenting processes occurring when the Linné ejecta formed. Nonetheless, given that multiple reference lunar studies (see the Discussion section) have systematically used the power-law curves to fit the boulders SFD, we prefer to use the power-law model to fit our data in order to compare the Linné obtained results with other similar analyses performed on the Moon.

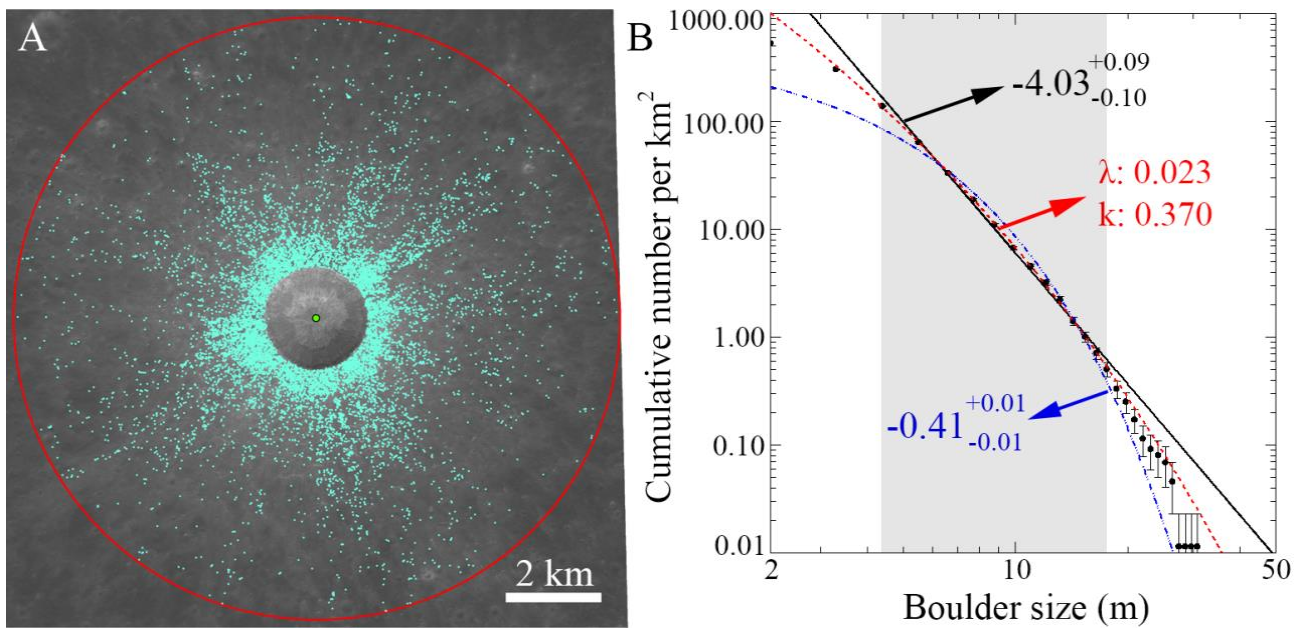

Fig. 6: A) The spatial distribution of the identified ejected boulders ≥ 4.4 m on the study area (124.01 km$^2$). B) SFD of the identified boulders (bin size 1.1 m, Table 2). Following Michikami et al., 2008, the vertical error bars indicate the root of the cumulative number of counted boulders divided by the considered area. The data were fitted between 4.4 and 17.6 m (highlighted in grey) with a power-law curve (solid black line), an exponential-law curve (dashed-point blue curve) and with a Weibull distribution with scale parameter $\lambda = 0.023$ and a shape parameter $k = 0.370$ (dashed red curve).

## 3.2.2 The radial boulder SFD analysis

In order to understand how the boulder sizes are radially distributed from the Linné's rim we prepared the plot of Fig. 7. As it is expected from impact cratering dynamics (Melosh 1989), the biggest sizes (> 15 m) are all located in close proximity to the crater's rim, while their frequency radially decreases with increasing distances from the rim. At distances >3.5 km from the Linné's rim the ejected boulders are all < 10 m in size.

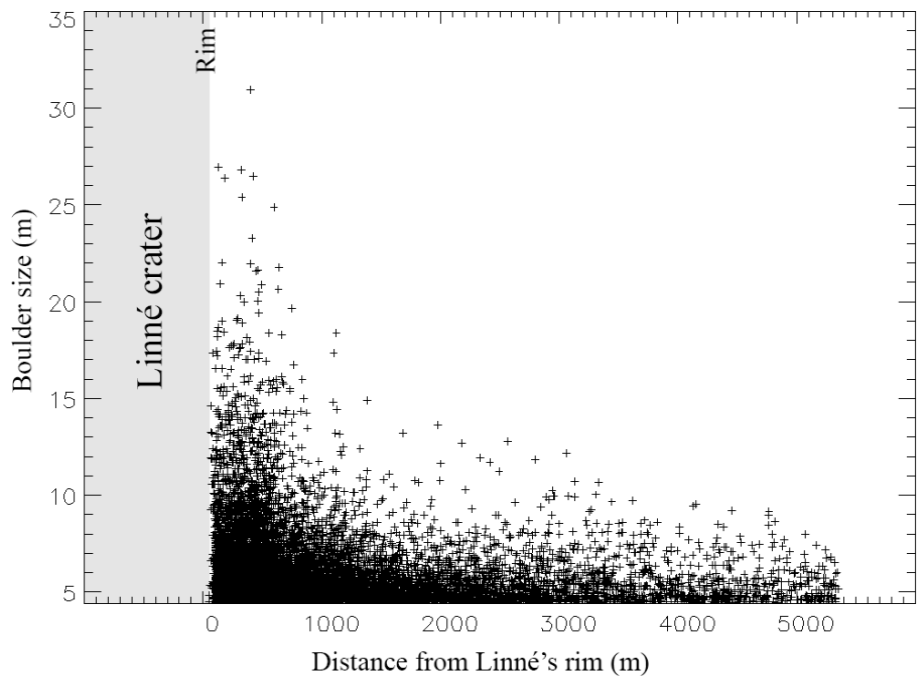

Fig. 7: A) The boulder size versus distance distribution surrounding the Linné's crater.

To quantitatively evaluate how the boulder SFD changes with the distance from the crater's rim, we selected circular rings with cumulatively increasing step sizes of 500 m, see Fig. 8A. We therefore extracted all the boulders falling inside such areas and computed the cumulative SFD with the same approach of section 3.2.1. We then fitted the power-law curves in the size ranges 4.4-17.6 m on all selections, as well as computed the density of boulders per $km^2$ both at size 4.4 m and 17.6 m, Fig. 8B. The results are presented in Table 2.

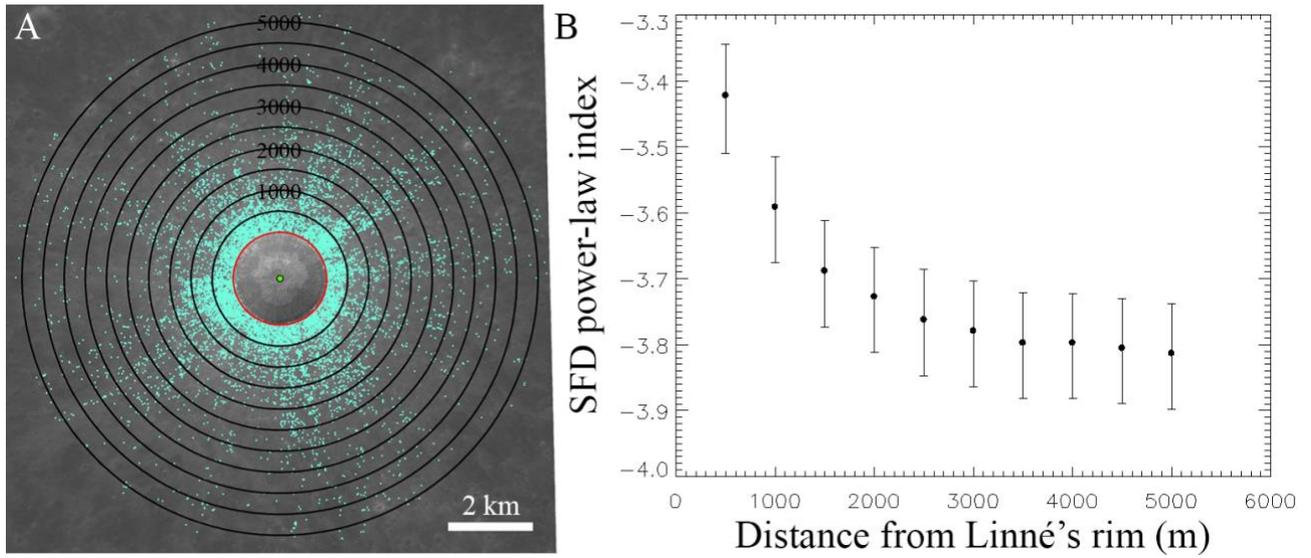

Fig. 8: A) The circular crowns used to compute the radial boulders SFD. B) The resulting boulder SFD power-law indices extracted from the circular ring of A).

Table 2: Cumulative boulder statistics as a function of the radial distance from the Linné's rim. The SFD power-law indices are all computed in the 4.4 – 17.6 m size range. The density of boulders per km$^2$ is also presented for two sizes.

| Distance from Linné's rim (m) | Power-law index | Errorbar + | Errorbar - | Density/km$^2$ @ 4.4m | Density/km$^2$ @ 17.6m |
|---|---|---|---|---|---|
| < 500 | -3.42 | 0.08 | 0.09 | 1428.64 | 13.45 |
| < 1000 | -3.59 | 0.08 | 0.09 | 848.63 | 6.07 |
| < 1500 | -3.69 | 0.08 | 0.09 | 569.92 | 3.56 |
| < 2000 | -3.73 | 0.08 | 0.09 | 407.45 | 2.39 |
| < 2500 | -3.76 | 0.08 | 0.09 | 307.22 | 1.71 |
| < 3000 | -3.78 | 0.08 | 0.09 | 240.20 | 1.31 |
| < 3500 | -3.80 | 0.08 | 0.09 | 192.49 | 1.02 |
| < 4000 | -3.80 | 0.08 | 0.09 | 157.60 | 0.84 |
| < 4500 | -3.81 | 0.08 | 0.08 | 131.68 | 0.69 |
| < 5000 | -3.82 | 0.08 | 0.09 | 111.45 | 0.58 |

### 3.2.3 The azimuthal boulder density analysis

In order to understand if there are any orientation anisotropies in the boulder densities, we computed the density map over cells with sizes of 50 m by 50 m over the entire study area. The resulting number of boulders per km$^2$ is indicated in Fig. 9, which shows that the greatest number of boulders per surface area are located within the first 2-3 km from the craters rim. Instead, at distances greater than 4 km from the rim it is possible to observe the ejected radial rays.

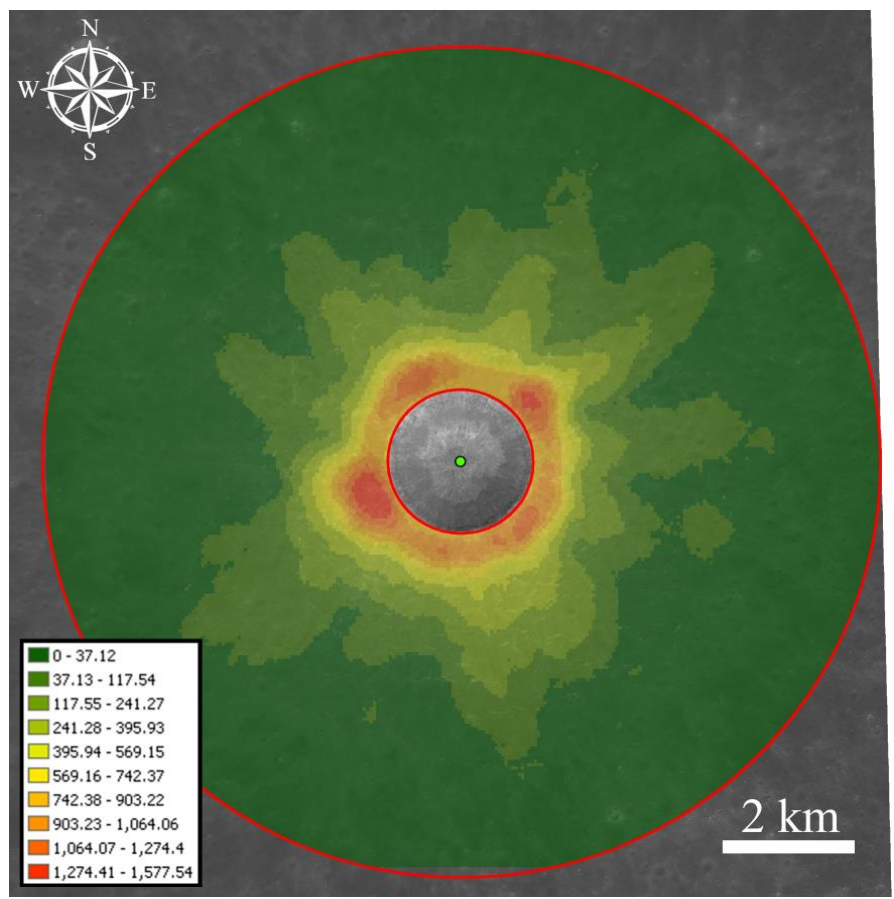

Fig. 9: Density map per km$^2$ of all boulders ≥ 4.4 m present in the study area.

To get a more quantitative measurement of the azimuthal inhomogeneity observed in Fig. 9 we selected 12 sectors surrounding the Linné crater, each one with an area of 10.33 km$^2$ and an angular dimension of 30° (starting from the North and oriented in the clockwise direction, see Fig. 10A, B).

For each sector we extracted the boulders SFD and computed the power-law indices, as well as the boulder densities. The resulting values are presented in Fig. 10C and in Table 3.

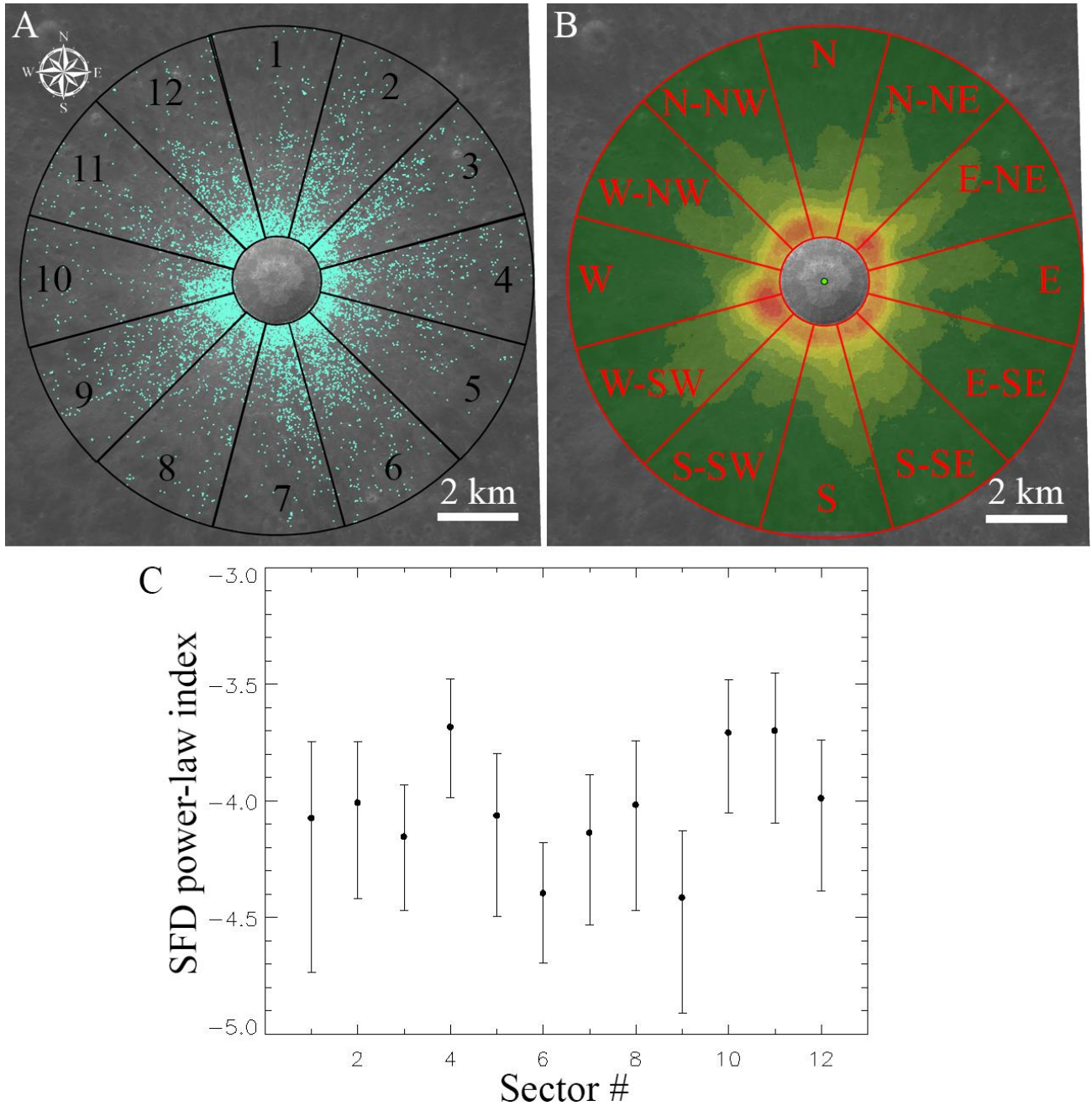

Fig. 10: A) The 12 sectors used to compute the radial boulders SFD. B) Same as A), but overlapped on the density map per km$^2$ of Fig. 9. C) The resulting boulder SFD power-law indices extracted from each sector #.

Table 3: Cumulative boulder statistics for the identified sector #. The available fitting ranges used to compute the power-law indices are indicated. The density of boulders per km$^2$ is also presented for two sizes.

| Sector # | Fitting range (m) | Power-law index | Errorbar + | Errorbar - | Density/km$^2$ @ 4.4m | Density/km$^2$ @ 17.6m |
|---|---|---|---|---|---|---|
| 1 | 4.4-17.6 | -4.07 | 0.33 | 0.66 | 69.99 | 0.19 |
| 2 | 4.4-15.4 | -4.01 | 0.26 | 0.41 | 85.38 | 0.48 |
| 3 | 4.4-15.4 | -4.15 | 0.22 | 0.32 | 118.97 | 0.19 |
| 4 | 4.4-17.6 | -3.68 | 0.21 | 0.30 | 80.35 | 0.68 |
| 5 | 4.4-15.4 | -4.06 | 0.27 | 0.43 | 77.44 | 0.34 |
| 6 | 4.4-13.2 | -4.40 | 0.22 | 0.30 | 114.62 | - |
| 7 | 4.4-17.6 | -4.14 | 0.25 | 0.40 | 108.23 | 0.39 |
| 8 | 4.4-17.6 | -4.02 | 0.28 | 0.45 | 83.45 | 0.39 |
| 9 | 4.4-17.6 | -4.42 | 0.29 | 0.50 | 142.69 | 0.29 |
| 10 | 4.4-17.6 | -3.71 | 0.23 | 0.34 | 86.93 | 0.48 |
| 11 | 4.4-17.6 | -3.70 | 0.25 | 0.40 | 78.22 | 0.39 |
| 12 | 4.4-17.6 | -3.99 | 0.25 | 0.40 | 115.00 | 0.39 |

## 4.0 Discussion

The identification through orbital imagery of the ejected boulder SFD surrounding lunar craters became systematic thanks to the NASA Lunar Orbiter images (Cintala and McBride, 1995). In particular, the debris field surrounding the Surveyor I, III, VI landing sites have been studied through high resolution imagery, identifying boulders with sizes ranging from 1 to 6.5 m and returning fitting power-law curves with indices from -3.51 ± 0.47 to -6.02 ± 1.40 (Table 4). For the specific case of the boulders surrounding the Surveyor VII landing sites the sizes 10-65 m have been investigated, returning a power-law index of -4.03 ± 0.14.

Afterwards, on the basis of NASA Lunar Orbiter III, V and the Apollo 17 images, Bart and Melosh (2010a) identified the boulders SFD of multiple craters derived through sizes ranging from 3 to 100s of metres. This dataset (Table 4) shows that most of the indices fall in the -3.0 to -4.5 range, with a minimum value of -2.2 and a maximum one of -5.5. Such power-law indices are considered to be all collisional impact-related given the neighboring presence of a craters.

In this paper, a manual detection of boulders surrounding the Linné crater was performed. In the analysis we considered the entire area with a radius of 6.4 km from the craters' centre. Through a manual procedure, we identified 12067 boulders ≥ 4.4 m deriving a global boulder SFD with a power-law index of -4.02 +0.09/-0.10. This value confirms the impact cratering formation power-law ranges and it is almost identical with the one obtained by Bart and Melosh (2010a) on the V-167-H2 crater (located on the Highlands terrain at 20.2°W, and 10.1°N) and on the V-211-H3 crater (located on a Mare terrain at 56.1°W, 13.7°N), 0.69 km and 0.52 km wide, respectively. Moreover, the size ranges used on the V-167-H2 and on the V-211-H3 craters to compute the -4.0 power-law indices fall in the 4.0-18.0 m range, comparable to the one we used (4.4-17.6 m). Basilevsky et al. (2013) indicated that on lunar craters with an age between 40 and 80 Ma, 50% of the original ejected boulder population ≥ 2 m disappears, while 99% of such boulders are destroyed in about 150-300 Ma. Given the proposed formation age of the Linné crater (< 10 Ma, Stickle et al., 2016) the surrounding ejecta and their size-frequency distribution can be considered pristine (Basilevsky et al., 2013).

The radial distribution analysis presented in Fig. 8 and Table 2, shows that as the distance from Linné's rim increases, the boulder SFD power-law index steepens, ranging from -3.42 in the first 500 m to -3.81 at distances 10 times larger. This is directly related to the impact cratering formation dynamics. Indeed, during an impact event, the biggest boulders are ejected closer to the newly formed rim, requiring much more kinetic energy with respect to the small chunks to travel at the same distance from the crater's centre (Melosh, 1989). Instead, the smallest chunks of material are present not only in close proximity to the rim, but also at much bigger distances. The relative abundance of the biggest sizes is much more prominent within the first km from the rim, hence resulting in a shallower power-law index than the more distal surroundings. Instead, at a much bigger distance the relative abundances of the small sizes become prominent with respect to the full statistics (as indicated in Fig. 7 too), hence steepening the SFD. This trend is confirmed for the size end-members (4.4 and 17.6 m) densities per km$^2$ (Table 2).

Table 4: The SFD derived for lunar boulders and the corresponding size range at which they were derived.

| Area/Crater | Mission | Reference | Power-law index | Diameter size/Range | Notes |
|---|---|---|---|---|---|
| Surveyor I landing site | NASA/Lunar Orbiter | Cintala and McBride (1995) | -3.51 ± 0.47 | 1 - 5 m | Field debris surrounding Surveyor I landing - impact related |
| Surveyor III landing site | NASA/Lunar Orbiter | Cintala and McBride (1995) | -5.65 ± 0.63 | 1 - 6.5 m | Field debris surrounding Surveyor III landing - impact related |
| Surveyor VI landing site | NASA/Lunar Orbiter | Cintala and McBride (1995) | -6.02 ± 1.40 | 1 - 4 m | Field debris surrounding Surveyor VI landing - impact related |
| Surveyor VII landing site | NASA/Lunar Orbiter | Cintala and McBride (1995) | -4.03 ± 0.14 | 10 - 65 m | Field debris surrounding Surveyor VII landing - impact related |
| Crater III-185-H3 | NASA/Lunar Orbiter III | Bart and Melosh (2010) | -4.2 | 3 - 10 m | Collisional Impact related to crater III-185-H3 |
| Crater III-168-H2 | NASA/Lunar Orbiter III | Bart and Melosh (2010) | -4.0 | 4 - 18 m | Collisional Impact related to crater III-168-H2 |
| Crater III-186-H3 | NASA/Lunar Orbiter III | Bart and Melosh (2010) | -4.0 | 3 - 18 m | Collisional Impact related to crater III-186-H3 |
| Crater III-189-H2 | NASA/Lunar Orbiter III | Bart and Melosh (2010) | -3.0 | 4 - 15 m | Collisional Impact related to crater III-189-H2 |
| Crater V-63-H2 | NASA/Lunar Orbiter V | Bart and Melosh (2010) | -3.0 | 10 - 70 m | Collisional Impact related to crater V-63-H2 |
| Crater V-82-M | NASA/Lunar Orbiter V | Bart and Melosh (2010) | -3.0 | 50 - 320 m | Collisional Impact related to crater V-82-M |
| Crater V-152-H2 | NASA/Lunar Orbiter V | Bart and Melosh (2010) | -5.5 | 5 - 18 m | Collisional Impact related to crater V-152-H2 |
| Crater V-153-H2 | NASA/Lunar Orbiter V | Bart and Melosh (2010) | -3.0 | 5 - 45 m | Collisional Impact related to crater V-153-H2 |
| Crater V-167-H2 | NASA/Lunar Orbiter V | Bart and Melosh (2010) | -4.0 | 4 - 18 m | Collisional Impact related to crater V-167-H2 |
| Crater V-167-H3 | NASA/Lunar Orbiter V | Bart and Melosh (2010) | -4.0 | 5 - 20 m | Collisional Impact related to crater V-167-H3 |
| Crater V-199-M | NASA/Lunar Orbiter V | Bart and Melosh (2010) | -2.2 | 50 - 380 m | Collisional Impact related to crater V-199-M |
| Crater V-211-H3 | NASA/Lunar Orbiter V | Bart and Melosh (2010) | -4.0 | 4 - 18 m | Collisional Impact related to crater V-211-H3 |
| Ap17-Pan-2345a | NASA/Apollo 17 | Bart and Melosh (2010) | -4.0 | 10 - 40 m | Collisional Impact related to crater Ap17-Pan-2345a |
| Ap17-Pan-2345b | NASA/Apollo 17 | Bart and Melosh (2010) | -4.5 | 10 - 30 m | Collisional Impact related to crater Ap17-Pan-2345b |
| Ap17-Pan-2345e | NASA/Apollo 17 | Bart and Melosh (2010) | -4.5 | 10 - 22 m | Collisional Impact related to crater Ap17-Pan-2345e |
| Ap17-Pan-2345f | NASA/Apollo 17 | Bart and Melosh (2010) | -4.0 | 10 - 30 m | Collisional Impact related to crater Ap17-Pan-2345f |
| Ap17-Pan-2345g | NASA/Apollo 17 | Bart and Melosh (2010) | -4.5 | 10 - 30 m | Collisional Impact related to crater Ap17-Pan-2345g |
| Ap17-Pan-2345i | NASA/Apollo 17 | Bart and Melosh (2010) | -3.2 | 10 - 50 m | Collisional Impact related to crater Ap17-Pan-2345i |
| Censorinus crater | NASA/LRO | Krishna and Senthil Kumar (2016) | -2.76 | 2-25 m | Collisional Oblique Impact related to Censorinus crater |
| Linné crater | NASA/LRO | This study | -4.03 +0.09/-0.10 | 4.4-17.6 m | Collisional Impact related related to Linné crater |

Much more complicated and difficult to interpret are the boulder distribution anisotropies observed in Fig. 9 and Fig. 10, and quantified in Table 3. By using the 2D iSALE shock physics code, Martellato et al. (2017) modelled the Linnè crater formation as a perpendicular impact of a projectile impacting the Moon with a velocity of 18 km s$^{-1}$. Such vertical impact was taken as an approximation because of the apparent symmetric pattern of Linnè ejecta blanket. Instead, from our ejecta distribution analysis (Fig. 9), we can observe that the boulder density is not perfectly symmetrical around the craters' centre, resulting in a slight excess of boulders' density in the NE-SW direction. This small anisotropy of the Linné proximal ejecta has been also observed during the mapping phase of this work, as shown in the geological map of Fig. 5, and may be the marker of an eventual projectile obliquity, even considering that the circularity of a final crater is relatively well-preserved down to an impact angle of 30° (e.g. Pierazzo and Melosh, 2000)[3]. This explanation has been proposed by Krishna and Senthil Kumar (2016) to support both the crater shape as well as the clear ejecta anisotropies surrounding the Censorinus crater (a 3.8 km crater located in the lunar highlands in close proximity to the Mare Tranquillitatis), even if in that case the asymmetrical topography is much more evident than in that of Linnè (Fig. 2).

Another equally possible explanation for the ejecta distribution of Fig. 9 may be related to an impact on a target characterised by the presence of different layer materials. Target rheological properties, such as internal strength, porosity, friction coefficient and layering, are known to strongly affect the final craters' morphology (e.g. Elbeshausen et al., 2009; Wünnemann et al., 2011). Martellato et al. (2017) tested several target configurations for the formation of Linnè crater, as the one modeling the target as a double basaltic layer where the upper one has a thickness varying between 50 and 400 m and is characterised by a higher degree of fracturation. Such layers approximated more recent magmatic flows emplaced over older solidified lava pulses that can differ in mechanical properties. This is also in agreement with the wall change in brightness located at about 200 m below the rim, that might point to a different material composition. Hence, a non-homogeneous impacted surface could also result in the slight asymmetrical boulder distribution we see. In addition, the boulder SFD power-law indices of Table 3 greatly vary between -3.68+0.21/-0.30 (sector #4) and -4.42+0.29/-0.50 (sector #9). Their different values confirm the SFD variability that can be found when different azimuth directions are considered (Krishna and Senthil Kumar, 2016). Nevertheless, if we consider the associated errorbars the power-law indices generally overlap, with the exception of sectors #4 (East), 9, 10 and 11 (West to North direction). Such SFD behaviour is not correlated with the NE-SW density values we observed, and it may be the direct result of the different target rheological properties that characterised the Linné site before the crater formation.

---

[3] The small asymmetry observed in the ejecta distribution suggests that the impact generating the Linné crater could be oblique. Nevertheless, an impact angle of 30° can only be considered as the lower limit given that this is the minimum impact angle within the circularity of the crater is ensured (Pierazzo and Melosh, 2000). To quantitatively evaluate the azimuth direction of the impactor multiple hydrocode simulations are needed, but they will require a future, dedicated effort.

The possibility of a non-homogeneous pre-impacted Linnè site is also supported by the presence of different ejected radial rays observed in Fig. 9 and 10. Indeed Sabuwala et al. (2018) recently indicated that radial streaks of ejecta surrounding the Moon and other planetary bodies' craters can form where the pre-impacted surface includes undulations and surface inhomogeneities.

We finally investigated the possible relation between the boulders ejecta distribution and the regolith thickness affecting the Linnè impact site, since local variations of regolith thickness may affect the boulder's travel distance. Indeed, Bart and Melosh (2010b) studied the effects of a regolith blanket on the ejecta emplacement for craters in the range of hundreds of metres, finding that the shallower the regolith layer is, the further the rocky blocks are ejected from the crater's rim. A 4 m thick regolith layer was found at Mare Serenitatis on the basis of Earth-based radar and optical data (Shkuratov and Bondarenko 2001). A mean thickness of 2-4 m in mare regions was also found from the analysis of crater geometry (Bart et al. 2011), while the same methodology applied to the Sinus Iridum region gave a regolith thickness ranging from about 5 to 11 m (Fa et al., 2015). These studies suggest that the regolith blanket at the Linnè impact site is likely a thin layer. A rough estimate of the regolith thickness at the Linnè impact site can be performed by using eq. 9 of Bart and Melosh (2010b):

$$f_{max} = \frac{138.5}{g}\left(\frac{D_c}{H}\right)^{0.7086},$$

where $f_{max}$ is the distance the boulder is thrown in m, $g$ is the lunar gravitational acceleration in m/s², $D_c$ is the crater diameter in m, and $H$ is the regolith depth in m.

The above equation allowed to compute the maximum distance of boulders ejected on a 45° ballistic trajectory, when the target is made up of a layer of regolith on top a consolidated substratum. In the case of a 4-m regolith layer, the expected maximum ejecta range is ~7.5 km for a Linnè sized crater, while it can reach 20 km if the regolith thickness drops down to 1 m. From our boulders size-frequency distribution coupled with the Linnè geological map, we find that very few boulders > 4.4 m are identified after 6.4 km. Hence, by using this 6.4 km limit, the corresponding regolith thickness of the Linnè impact site should be ~4.75 m, strongly supporting the regolith thicknesses previously indicated in Mare Serenitatis.

## 5.0 Summary and Conclusions-------------------------------------------------------------

Through manual detection on high-resolution LROC images we identified 12067 ejected boulders ≥ 4.4 m surrounding the Linné crater, a simple crater with an age < 10 ma (Stickle et al., 2016) located in northwestern Mare Serenitatis. The resulting SFD derived in the range 4.4 – 17.6 m is fitted with a power-law curve with an index of -4.03 +0.09/-0.10. This value falls within the impact cratering formation power-law ranges identified on other lunar craters (Cintala and McBride, 1995; Bart and Melosh 2010a) and supports the interpretation that all boulders surrounding the Linné crater are ejected blocks formed during the crater emplacement.

By studying the radial ejecta abundances, we found that, as the distance from Linné's rim increases, the boulder SFD power-law index steepens, ranging from -3.42 in the first 500 m to -3.82 at distances 10 times larger. This means that the relative abundance of the biggest sizes is largest within the first km from the rim (as expected from impact cratering dynamics; Melosh, 1989), hence resulting in a shallower power-law index. On the other hand, at bigger distances the relative abundances of the smaller boulder sizes increases, becoming dominant with respect to the full statistics and hence steepening the SFD.

The Linné high-resolution geological map showed that its proximal ejecta blanket is slightly asymmetrical in the NE-SW direction. This is confirmed by our boulder surface density analysis. Such behavior can be either the result of an oblique impact emplacement of the original impactor, that ejected more boulders with this preferential distribution, or the result of a perpendicular impact in the Mare Serenitatis location, but on a surface characterised by lunar basalts with different local mechanical properties. We suggest that both scenarios can be equally possible to explain the ejecta distribution anisotropies we observe.

Eventually, we investigated the possible relation between the boulders ejecta distribution and the regolith thickness affecting the Linnè impact site, since local variations of regolith thickness may affect the boulder's travel distance (Bart and Melosh, 2010b). By exploiting our boulders statistics and our Linnè geological map, coupled with Eq. 9 of Bart and Melosh (2010b), we estimated that the corresponding regolith thickness of the Linnè impact site should be ~4.75 m, hence supporting the previously indicated Mare Serenitatis' regolith thickness (Shkuratov and Bondarenko, 2001).

**Acknowledgements**

We would like to thank the two anonymous referees for constructive comments that lead to a great improvement of the paper. This Paper is part of a project that has received funding from the European Union's Horizon 2020 research and innovation programme under grant agreement Nº776276 (PLANMAP). We gratefully acknowledge the NASA LROC Archive node for providing access to the LROC dataset used in this work (http://wms.lroc.asu.edu/lroc/). We made use of the ARCGIS 10.1 software and IDL to perform the presented analysis. We thank Dr. Valerio Vivaldi for important discussions about the Linné formation and its geological setting. The boulder statistics presented in this paper are available upon request sent to the two email addresses: maurizio.pajola@inaf.it and maurizio.pajola@gmail.com.

**References**

Bart, G. D. and Melosh, H. J. (2010a). Distributions of boulders ejected from lunar craters, Icarus, 209, pp. 337-357.

Bart, G. D. and Melosh, H. J. (2010b). Impact into lunar regolith inhibits high-velocity ejection of large blocks. Journal of Geophysical Research, Volume 115, Issue E8.

Bart, G.D., Nickerson, R.D., Lawder, M.T., and Melosh H.J. (2011). Global survey of lunar regolith depths from LROC images. Icarus, 215, 485-490.

Basilevsky, A.T., Head, J.W., and Horz, F. (2013). Survival times of meter-sized boulders on the surface of the Moon, Planetary and Space Science, 89, 118-126.

Binder, A. B., Arvidson, R. E., Guinness, E. A., Jones, K. L., Mutch, T. A., Morris, E. C., Pieri, D. C., and C. Sagan (1977). The geology of the Viking Lander 1 site. Journal of Geophysical Research., 82, pp. 4439–4451.

Brown, W. K., and Wohletz, K. H. (1995). Derivation of the Weibull distribution based on physical principles and its connection to the Rosin-Rammler and lognormal distributions. J. Appl. Phys., 78, 2758-2763.

Christensen, P. R. (1986). The spatial distribution of rocks on Mars. Icarus, 68, 217-238, http://dx.doi.org/10.1016/0019-1035(86)90020-5

Cintala, M. J., McBride, K. M. (1995). Block distributions on the lunar surface: a comparison between measurements obtained from surface and orbital photography. NASA Technical Memorandum 104804. This publication is available from the NASA Centre for AeroSpace Information, 800 Ellaidge Landing Road, Linthicum Heights, MD 21090-2934, (301) 621-0390.

Craddock, R. A., Golombek, M., and Howard, A. D. (2000) "Analyses of rock size-frequency distributions and morphometry of modified Hawaiian lava flows: Implications for future Martian landing sites" Lunar and Planetary Science XXXI, Abstract #1649, Lunar and Planetary Institute, Houston (CD-ROM).

Elbeshausen, D., Wünnemann, K., and Collins, G.S. (2009). Scaling of oblique impacts in frictional targets: Implications for crater size and formation mechanisms. Icarus, 204, 716–731.

Fa, W., Zhu, M.-H., Liu, T., and Plescia, J.B. (2015). Regolith stratigraphy at the Chang'E-3 landing site as seen by lunar penetrating radar. Geophysical Research Letters 42, 10,179-10,187.

Garvin, J. B., Mouginis-Mark, P. J., and Head, J. W. 1981). Characterization of rock populations on planetary surfaces: Techniques and a preliminary analysis of Mars and Venus. Moon and Planets 24, 355 – 387.

Garvin, J.B., Robinson, M.S., Frawley, J. et al. (2011). Linné: Simple lunar mare crater geometry from LRO observations. 42 LPSC, 2063.

Geissler, P., Petit, J.-M., Durda, D. D. et al. (1996), Erosion and Ejecta Reaccretion on 243 Ida and Its Moon, Icarus, 120, 140.

Golombek, M., and D. Rapp (1997). Size-frequency distributions of rocks on Mars and Earth analog sites: Implications for future landed missions. Journal of Geophysical Research, 102, 4117-4129. http://dx.doi.org/10.1029/96JE03319

Golombek, M. P. et al. (2003). Rock size-frequency distributions on Mars and implications for Mars Exploration Rover landing safety and operations. Journal of Geophysical Research, 27, 8086 http://dx.doi.org/10.1029/2002JE002035

Golombek, M. P. et al. (2008). Size-frequency distributions of rocks on the northern plains of Mars with special reference to Phoenix landing surfaces. Journal of Geophysical Research, 113, E00A09 http://dx.doi.org/10.1029/2007JE003065

Grady, D. E., and Kipp, M. E. (1987). Dynamic rock fragmentation. In Fracture Mechanics of Rock, edited by B. K. Atkinson, pp. 429-475, Academic Press, London, U.K.

Grant, J. A. et al. (2006). Distribution of rocks on the Gusev Plains and on Husband Hill, Mars. Geophysical Research Letters, 33, L16202 http://dx.doi.org/10.1029/2006GL026964

Hartmann, W. K. (1969). Terrestrial, Lunar, and Interplanetary Rock Fragmentation, Icarus, 10, 201.

Hiesinger H., Jaumann R., Neukum G., and Head J. W. (2000). Ages of mare basalts on the lunar nearside. Journal of Geophysical Research: Planets, 105:29,239–29,275.

Klem, S. M., Henriksen, M. R., Stopar, et al. (2014), Controlled LROC Narrow Angle Camera High Resolution Mosaics. In *Lunar and Planetary Science Conference* (Vol. 45, abstract #2885).

Krishna, N. and Senthil Kumar, P. (2016). Impact spallation processes on the Moon: A case study from the size and shape analysis of ejecta boulders and secondary craters of Censorinus crater. Icarus 264, 274-299.

Küppers, M., Moissl, R., Vincent, J.-B., et al. (2012). Boulders on Lutetia, Planetary and Space Science, 66, 71.

Kuiper, G. P. (1965). The Surface Structure of the Moon. In *The Nature of the Lunar Surface,* pp. 99–105. The Johns Hopkins Press, Baltimore.

Martens, H. R., Ingersoll, A. P., Ewald, S. P., et al. (2015). Spatial distribution of ice blocks on Encelkadus and implications for their origin and emplacement. Icarus, 245, 162-176.

Martellato. E., Vivaldi, V., Massironi, M., et al. (2017). Is the Linné impact crater morphology influenced by the rheological layering on the Moon's surface? Insights from numerical modeling. Meteoritics & Planetary Science, 1-24.

Mazrouei, S., Daly, M. G., Barnouin, O. S., Ernst, C. M., & DeSouza, I. (2014). Block distributions on Itokawa. Icarus, 229, 181.

McGetchin, T. R., M. Settle, and J. W. Head (1973), Radial thickness variation in impact crater ejecta: Implications for lunar basin deposits, Earth Planet Sci. Lett., 20, 226–236.

Melosh, H. J. (1989), Impact Cratering, 245 pp., Oxford Univ. Press, New York.

Michikami, T., Nakamura, A. M., Hirata, N., et al. (2008). Size-frequency statistics of boulders on global surface of asteroid 25143 Itokawa. *Earth, Planets and Space*, Volume 60, p. 13-20.

Moore, H. J. and Jakosky, B. M. (1989). Viking landing sites, remote-sensing observations, and physical properties of Martian surface materials. Icarus, 81, 164-184.

Mottola, S., Arnold, G., Grothues, H.-G., et al. (2015). The structure of the regolith on 67P/Churyumov-Gerasimenko from ROLIS descent imaging. Science, 349, Issue 6247, aab0232.

Nyquist, H. (1928). Certain topics in telegraph transmission theory. Trans. Am. Inst. Elect. Eng., 47, 617.

Pajola, M., Vincent, J. B., Guettler, C. et al. (2015). Size-frequency distribution of boulders ≥ 7 m on comet 67P/Churyumov-Gerasimenko. Astronomy and Astrophysics, 583, A37.

Pajola, M., Lucchetti, A., Bertini, I. et al., (2016a). Size-frequency distribution of boulders ≥ 10 m on comet 103P/Hartley 2. Astronomy and Astrophysics, 585, A85.

Pajola, M., Lucchetti, A., Vincent, J. B. et al., (2016b). The southern hemisphere of 67P/Churyumov-Gerasimenko: Analysis of the preperihelion size-frequency distribution of boulders ≥ 7 m. Astronomy and Astrophysics, 592, L2.

Pajola, M., Oklay, N., La Forgia, F. et al., (2016c). Aswan site on comet 67P/Churyumov-Gerasimenko: Morphology, boulder evolution, and spectrophotometry. Astronomy and Astrophysics, 592, A69.

Pajola, M., Mottola, S., Hamm, M. et al., (2016d). The Agilkia boulders/pebbles size-frequency distributions: OSIRIS and ROLIS joint observations of 67P surface. MNRAS doi:10.1093/mnras/stw2720.

Pajola, M., Höfner, S., Vincent, J.B., et al. (2017a). The pristine interior of comet 67P revealed by the combined Aswan outburst and cliff collapse. Nature Astronomy, 1, 0092.

Pajola M., Rossato, S., Baratti, E. et al. (2017b). Boulders abundances and size-frequency distributions on Oxia Planum-Mars: Scientific implications for the 2020 ESA ExoMars rover. Icarus, 296, 73-90.

Pierazzo E. and Melosh H. J. (2000). Understanding oblique impacts from experiments, observations, and modeling. Annual Review of Earth and Planetary Sciences 28:141–167.

Pommerol A., Kofman W., Audouard J., et al. (2010). Detectability of subsurface interfaces in lunar maria by the LRS/SELENE sounding radar: Influence of mineralogical composition. Geophysical Research Letters 37:L03201.

Sabuwala, T., Butcher, C., Gioia, G. and Chakraborty, P. (2018). Ray systems in granular cratering. Physical Review Letters, 120, 264501.

Shkuratov Y.G. and Bondarenko N.V., (2001). Regolith Layer Thickness Mapping of the Moon by Radar and Optical Data. Icarus 149, 329-338.

Senthil Kumar, P., K. J. Prasanna Lakshmi, N. Krishna, R. Menon, U. Sruthi, V. Keerthi, A. Senthil Kumar, D. Mysaiah, T. Seshunarayana, and M. K. Sen (2014). Impact fragmentation of Lonar Crater, India: Implications for impact cratering processes in basalt, *J. Geophys. Res. Planets*, 119, 2029–2059, doi:10.1002/2013JE004543.

Shoemaker, E. M. et al. (1970). Apollo 12: Preliminary Science Report, NASA SP-235.

Shoemaker, E.M. (1987). Meteor Crater, Arizona, Geological Society of America Centennial Field Guide - Rocky Mountain Section.

Stickle A. M., Patterson G. W., Cahill J. T. S., and Bussey D. B. J. (2016). Mini-RF and LROC observations of mare crater layering relationships. Icarus 273:224–236.

Thomas, P. C., Veverka, J., Robinson, M. S., and Murchie, S. (2001). Shoemaker crater as the source of most ejecta blocks on the asteroid 433 Eros, Nature, 413, 394.

Turcotte, D. L. (1997). Fractals and chaos in geology and geophysics. Second Edition ed. Cambridge University Press, Cambridge.

Ward, J. G. et al. (2005). The size-frequency and areal distribution of rock clasts at the Spirit landing site, Gusev Crater, Mars. Geophysical Research Letters 32, L11203 http://dx.doi.org/10.1029/2005GL022705

Yingst, R. A., Haldemann, A. F. C., Biedermann, K. L., and Monhead, A. M. (2007). Quantitative morphology of rocks at the Mars Pathfinder landing site. Journal of Geophysical Research 112, E06002 http://dx.doi.org/10.1029/2005JE002582

Weider S. Z., Crawford I. A., and Joy K. H. (2010). Individual lava flow thicknesses in Oceanus Procellarum and Mare Serenitatis determined from Clementine multispectral data. Icarus, 209, 323–336.

Wentworth C. K., (1922). A Scale of grade and class terms for clastic sediments. J. Geol., 30, 377, pp. 377-392.

Wünnemann K., Nowka D., Collins G. S., Elbeshausen D., and Bierhaus M. (2011). Scaling of impact crater formation on planetary surfaces: Insights from numerical modeling. In Proceedings of the 11th hypervelocity impact symposium, edited by Schäfer F. and Hiermaier S. Schriftenreihe Forschungsergebnisse aus der Kurzzeitdynamik, vol. 20. Freiburg: Fraunhofer EMI. pp. 1–16.